\RequirePackage{lineno}
\documentclass[12pt,
reprint,
showpacs,
amsmath,amssymb,aps,
]{revtex4-1}

\usepackage{amsfonts}    
\usepackage{graphicx}   
\usepackage{color}      
\usepackage{float}
\usepackage{fullpage}
\usepackage{bm}
\usepackage{subfig}
\usepackage{epstopdf}

\let\oldalign\align
\let\endoldalign\endalign
\renewenvironment{align}{%
  \linenomath
  \oldalign
}{%
  \endoldalign
  \endlinenomath
}

\let\oldequation\equation
\let\endoldequation\endequation
\renewenvironment{equation}{%
  \linenomath
  \oldequation
}{%
  \endoldequation
  \endlinenomath
}

\def\eeq{\relax}
\def\beq#1#2\eeq{\begin{equation}\label{#1}#2\end{equation}}
\def\bal#1#2\eal{\begin{align}\label{#1}#2\end{align}}
\def\bse#1#2\ese{\begin{subequations}\label{#1}#2\end{subequations}}
\newcommand{\ca}{\begin{cases}}
\newcommand{\ac}{\end{cases}}
\newcommand{\ma}{\begin{pmatrix}}
\newcommand{\am}{\end{pmatrix}}
\def\dd{\operatorname{d}} 

\newcommand{\er}\eqref

\def\ann#1{} 

\begin{document}

\title{A high transmission broadband gradient index lens \\    
using elastic shell acoustic metamaterial elements}
\author{Alexey S. Titovich}
\affiliation{Mechanical and Aerospace Engineering, Rutgers University, Piscataway, NJ 08854}
\author{Michael R. Haberman}
\affiliation{Applied Research Laboratories and Department of Mechanical Engineering, The University of Texas at Austin, Austin, TX  78758}
\author{Andrew N. Norris}
\email{norris@rutgers.edu}
\affiliation{Mechanical and Aerospace Engineering, Rutgers University, Piscataway, NJ 08854}




\begin{abstract}

  The use of cylindrical elastic shells as elements in acoustic metamaterial devices is  demonstrated through simulations and underwater measurements of  a cylindrical-to-plane wave lens.  Transformation acoustics (TA) of a circular region to a square dictates that the effective density in the lens remain constant and equal to that of water.  Piecewise approximation to the desired effective compressibility is achieved using a square array with elements based on the elastic shell metamaterial concept developed in  \cite{Titovich14a}.  The size of the elements are chosen based on availability of shells, minimizing fabrication difficulties. The tested device is neutrally buoyant 
	comprising  48 elements of nine different types of commercial shells made from  aluminum, brass, copper, and  polymers.  Simulations  indicate a  broadband range in which the device  acts as a cylindrical to plane wave lens. 
	The experimental findings confirm  the broadband quadropolar response  from approximately 20 to 40 kHz, with positive gain of the radiation pattern in the four plane wave directions.   
	
\end{abstract}

\maketitle

\section{Introduction}   \label{sec1}

The    distortion of sound waves in  materials with spatially varying index of refraction    can be controlled using concepts from rays and high frequency propagation.  For instance,  all ray paths     can be made to behave according to a prescribed pattern, e.g. convergence at a point on the other side.  Transformation acoustics (TA) goes further  in making  the material  replicate an equivalent volume of "virtual" acoustic fluid which faithfully mimics the wave equation itself rather than some asymptotic approximation. Such TA-based gradient index (GRIN) lenses fall under the umbrella of acoustic metamaterials, a field which has seen tremendous innovation in recent years.  However, to build a TA-GRIN lens in the laboratory often demands  compromise between the frequency range of operation, transmission loss and lensing effectiveness, particularly in water, the acoustic medium of interest here.

A successful TA-GRIN lens  simultaneously displaying  high transmission and accurate wave steering  can be achieved in water using a sonic crystal (SC)  array of elastic scatterers.   Quasi-periodic SCs are capable of filtering, guiding and/or steering an incident wave based on a gradient of effective properties~\cite{Chen01b,Miyashita05,Cai07c,Lin09,Lin09a,Climente10,Martin10,Romero-Garcia13,Lin12}. Unlike phononic crystals, SCs cannot support shear waves in the bulk, hence, energy loss to mode conversion is minimized.  
The localized effective acoustic properties of a SC element are an average of the fluid and contained elastic scatterer. These depend on the shape, filling fraction, the effective bulk modulus and the effective density of the scatterer. 
In order to display the inhomogeneity required for a GRIN lens, the properties of the elements have to differ in a quasi-continuous manner. This has been 
successfully achieved in air \cite{Climente10} and in water \cite{Martin10}   by fixing the lattice constant and varying the filling fraction of solid cylinder scatterers in the fluid unit cell. 
For  air-based SCs the cylinders can be modeled as rigid~\cite{Sanchez-Perez98,DT,Romero-Garcia13}.  For water-based SCs the elasticity of the scatterer is not only non-negligible, but essential in the modeling of such structures. 

\begin{figure}[h!]    
\includegraphics[width=3.4in]{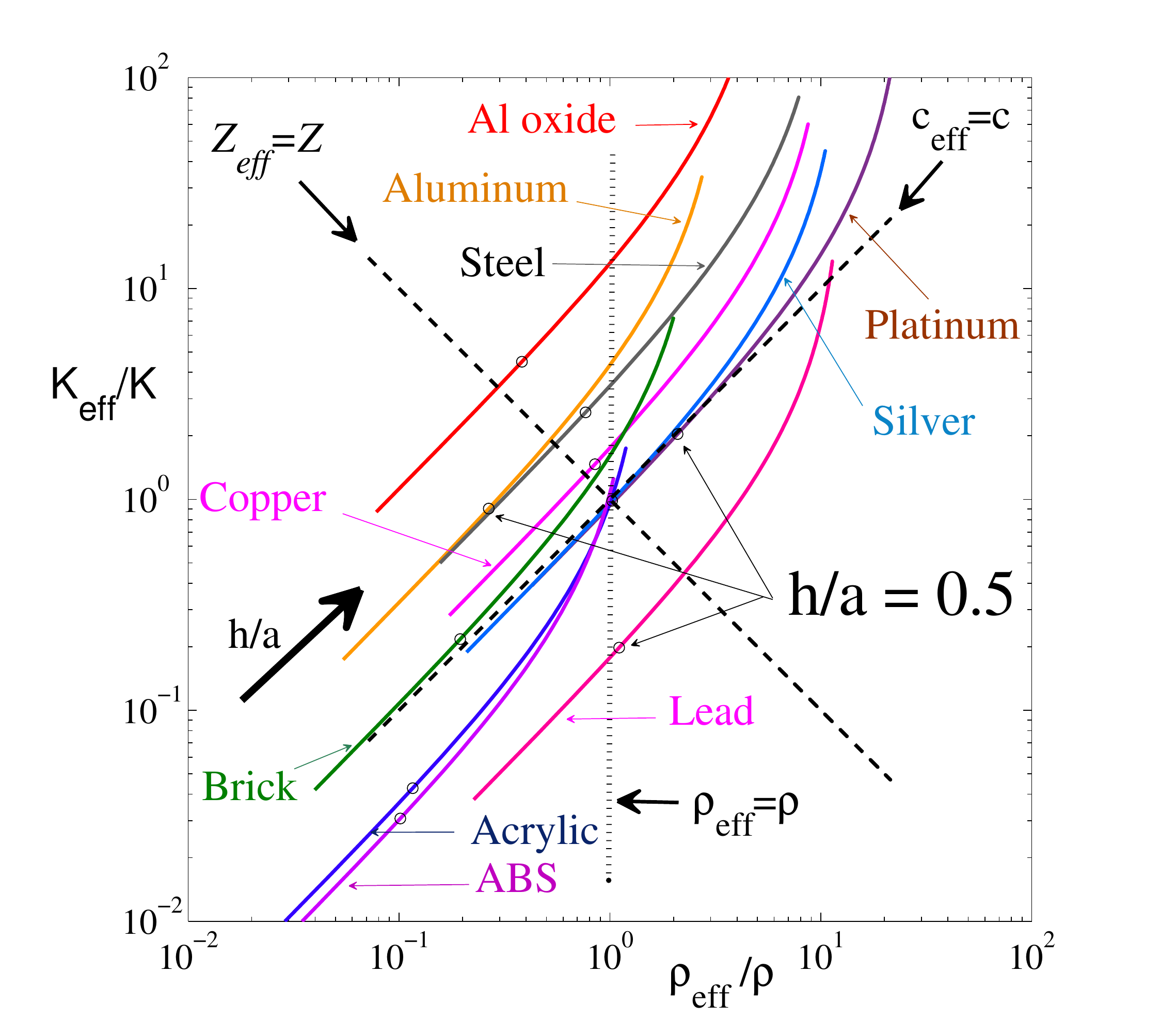}  
\caption{The effective density $\rho_\text{eff}$ and bulk modulus   $K_\text{eff}$  of hollow cylindrical shells for ten commonly available materials normalized relative to water, from eq.\ \eqref{-45}.  Each curve shows the properties as a function of the relative thickness to radius ratio $\frac ha$,  from small to large as indicated by the arrow.  Circles indicate the values for $\frac ha= 0.5$.   Diagonal dashed lines indicate  where the effective acoustic speed and impedance  coincide with those of water. (Color online)} 
\label{f1} 
\end{figure}

Typical engineering materials, such as metals, are much denser than water leading to  impedance mismatch and undesired scattering. A solution is to use a hollow air-filled elastic shell which has an effective density and bulk modulus much closer to that of water as compared to the solid material. The effective acoustic properties (speed, impedance)  depend  on the material of the shell and, in particular, on  its thickness \cite{Titovich14b, Titovich14a}.    The sensitivity of the effective compressibility to shell thickness is a consequence of hoop stress  in thin shells, which combined with the dependence of the effective density, results in the fact that thin shells have effective sound speed that is independent of thickness \cite{Titovich14a}.  The effective impedance, on the other hand, is a linear function of thickness in the same thin shell approximation.  These two basic facts together indicate that by choosing the material and the thickness, it is possible to achieve  a wide range of effective properties, as illustrated in the chart in Figure~\ref{f1} (motivated by earlier work in \cite{Martin12b}).   This is the central idea in the present work. 
  

The purpose of the present paper is to demonstrate the potential for TA-based GRIN lens design in water using the wide variety of shells available.  
The transformation acoustics example considered in detail here is the cylindrical-to-plane wave lens discussed  by Layman et al.~\cite{Layman11} It works by steering waves from a monopole source at the center away from the corners to the faces of the lens. The SC of Ref.~\cite{Layman11} is based on constructive multiple scattering from finite embedded elastic materials in a fluid matrix, something previously investigated by Torrent and Sanchez-Dehesa~\cite{Torrent07}. The GRIN lens device considered here expands the possibilities in Ref.~\cite{Torrent07} by increasing the range of achievable properties over those presented by Martin et al.~\cite{Martin12b}.  

The  cylindrical-to-plane wave lens is designed to increase  radiation in specific directions.
Enhanced directionality   has also been    experimentally observed \cite{Ke2007} 
for an acoustic source  placed inside a two-dimensional square lattice  phononic crystal operating at the band-edge frequency \cite{Qiu2006}.  
Highly directional  acoustic wave radiation is also  possible in  2D PCs at 
	 pass band frequencies far away from the band edge states, as shown in simulations of 
		 a square lattice of steel cylinders in water	\cite{Wen2009}. 
The use of the band structure of a periodic square array to produce  directional
water wave radiation was proposed by \cite{Mei2010}, and subsequently 
  demonstrated in  experimental measurements on a 6$\times$6 array  of surface-breaking cylinders \cite{Chekroun2015} with a monopolar source at the array center.  
Directional radiation has  been demonstrated in air using a non-periodic array of cylinders to produce scattering enhancement in the forward direction \cite{HaSkansson2007}.  Martin et al.~\cite{Martin2015,Rohde2015} produced acoustic GRIN focusing by changing the lattice constant in a PC with elastic shell elements.  Parallel zigzag rigid screens have also been proposed as potential focusing and directional beaming devices \cite{Tang2015}. 
While the spatial filtering  device described in this paper uses a fluid matrix,   Morval et al.\ \cite{Morvan2014}  show  directional enhancement  of a monochromatic acoustic  source into a surrounding water medium using a square array of cylinders in a solid matrix; the  2-dimensional   quadropolar collimation effect  is based on square-shaped equifrequency contours of the phononic crystal \cite{Vasseur2012}.  Although the solid matrix has obvious practical advantage, the narrow frequency device of \cite{Morvan2014} yields  decreased amplitude  in the preferential directions as compared with the free field radiation.   The TA-based device described here   does not have these limitations, and shows for the first time as far as we are aware, broadband positive gain in a neutrally buoyant square GRIN lens, with obvious implications for low loss underwater application.


The outline of the paper is as follows.  Transformation acoustics and the mapping for the cylinder-to-square lens are described in Section \ref{sec2}.   Acoustical properties of cylindrical  shells are discussed in Section \ref{sec3} and the proposed design using available cylindrical tubes is presented.  The experimental setup is described and acoustical measurements are discussed  in Section \ref{sec4}, with concluding remarks in Section \ref{sec5}.

\section{Conformal  transformation acoustics} \label{sec2}

The transformation of a circular region  to a square one can be achieved using a conformal change of coordinates.  Conformal mapping is a special case of the general theory of transformation acoustics (TA).  Usually, in TA one can expect the  material properties associated with a spatial transformation to display  anisotropy.  This could  be in the  density  \cite{Cummer07} or the bulk modulus  \cite{Norris08b}, or in both simultaneously   \cite{Norris09}, but usually something has to become anisotropic.  Conformal maps are unique in TA in that they do not require anisotropy.  In this case both the inertial  \cite{Cummer07} and the pentamodal  \cite{Norris08b} forms of TA converge, and there is no ambiguity or degrees of freedom, a feature that distinguishes  TA from its electromagnetic counterpart.  At the same time, there is some confusion in the application of TA for conformal mappings, e.g.\  \cite{Ren10}, so we briefly  review the correct procedure  \cite{Norris12a}.  

We are concerned with a background fluid (water) of density  $\rho$ and bulk modulus $K$ in which the acoustic pressure $p({\bf x})$ satisfies 
\begin{equation}\label{1}
\nabla ^2 p + \frac{\omega^2}{c^2}p=0,
\end{equation}
where $c = \sqrt{K/\rho}$ is the speed of sound and time harmonic dependence  $e^{-i \omega t}$  is understood. 
Under a conformal transformation $z \equiv x+iy  \to z_1(z) \equiv x_1+iy_1 $ the Laplacian $\nabla ^2 $ in the original variables becomes $|\dd z/\dd z_1|^2|\nabla_1 ^2 $.  If we define the pressure as   $p_1({\bf x}_1) = p({\bf x})$ then   $p_1$ satisfies the  Helmholtz equation in the mapped coordinates 
 with transformed acoustic speed 
$c_1 = | z_1'| c $ where $z_1' (z) = \dd z_1/\dd z$.  This means that the transformed parameters are indeed isotropic,  but it   does not provide unique expressions for the individual parameters $K_1$ and $\rho_1$, only the combination   $K_1/\rho_1 = c_1^2$.   The necessary second relation comes from the requirement that the pressure in the transformed fluid arises from a particle displacement field ${\bf u}_1({\bf x}_1)$ which satisfies    
the momentum equation 
$- \omega^2 \rho_1{\bf u}_1 = - \nabla_1 p_1 $ and the pressure constitutive relation
$p_1= -K_1\nabla_1\cdot {\bf u}_1$. 
Eliminating ${\bf u}_1$ gives the transformed Helmholtz equation for $p_1({\bf x}_1)$
if and only if $\rho_1$ is constant, which can be assumed equal to the original density.  In summary, the transformed parameters are 
\beq{3}
\rho_1 = \rho, \ \ \ K_1 = | z_1' (z)|^2 K. 
\eeq

\begin{figure} [h!]
 \includegraphics[width=3.in]{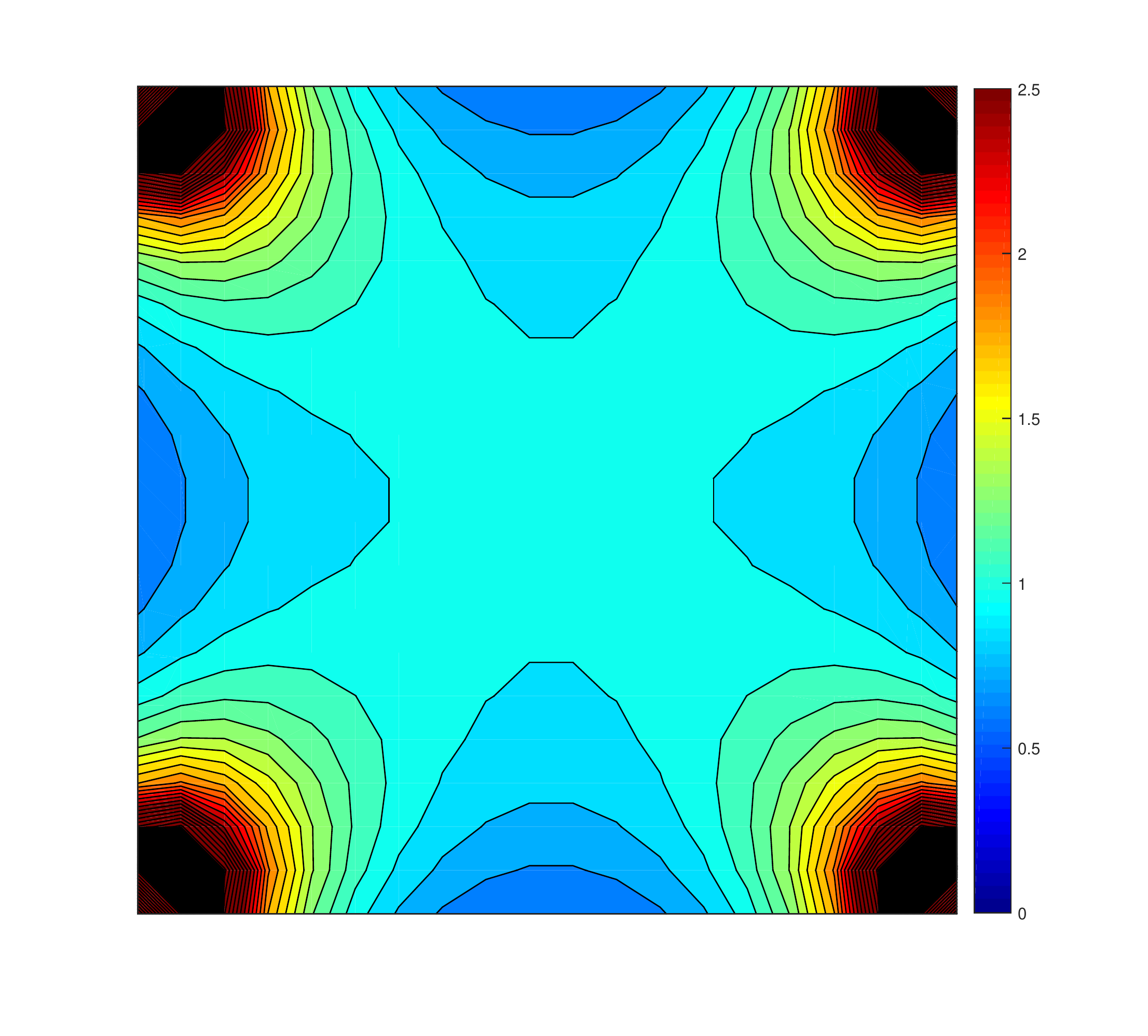} 
\caption{The  bulk modulus distribution $K_{1}/K$ for the cylindrical-to-square mapping.  (Color online)}  \label{f2}
\end{figure}

The lens is based the transformation of a circle of diameter $2b$ into a square of side $2b$, with the precise form of the circle-to-square mapping  given in the Appendix. In particular, we note from eqs.\  \eqref{3} and \eqref{1-15a}
that the mapped value of the bulk modulus associated with the original point $(r,\theta)$ in the circle is 
\beq{4}
K_1 = \frac{1.1636\, K}{\sqrt{\big(\frac rb\big)^8  +2\big(\frac rb\big)^4\cos 4\theta + 1}}.
\eeq
Along the principal directions $(\cos 4\theta =1)$ the bulk modulus decreases from  the center of the square to a global minimum  at the center of the sides.  Along the diagonals $(\cos 4\theta =-1)$
it increases from  its value at the center as it  becomes unbounded  at the four corners of the square.   The overall trend is illustrated in Figure \ref{f2}.

\section{Realization with  solid  shells as metamaterial elements} \label{sec3}

\subsection{Acoustical properties of cylindrical   shells}

\begin{figure}[h!]    
\centering
 \includegraphics[width=2.6in]{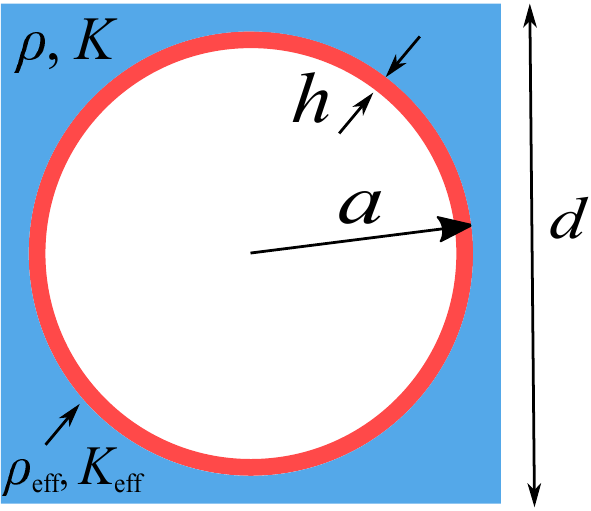}    
\caption{A square unit cell of a fluid saturated array of shells.  (Color online)}
\label{f3} 
\end{figure}

Consider a cylindrical shell of thickness $h$ and outer radius $a$  made of uniform solid with density $\rho_s$, shear modulus $\mu_s$, and Poisson's ratio $\nu_s$. The interior is air filled, which in the context of water as the ambient medium in the exterior means that we can safely ignore the inertia and stiffness of the interior.   
The shell's effective density is the average value taken over the circular region of radius $a$. 
The effective  bulk modulus is the value for which  the  radial compression of a uniform circular region of fluid under external pressure is the same as that of the shell under the same pressure, which follows  from plane strain elasticity \cite[p.\ 6]{Titovich2015c}.
In summary, 
\beq{-45}
 \begin{aligned}
\rho_\text{eff} &= \big(2h/a-(h/a)^2\big)\rho_s , 
\\
K_\text{eff}&=  \mu_s / \big(2(1-\nu_s) {\rho_s}/{\rho_\text{eff}} - 1\big). 
\end{aligned}
\eeq
The unit cell of the square array, shown in Figure~\ref{f3}, consists of a solid cylindrical shell surrounded by a complementary region of water.
The equivalent density and bulk modulus, $\rho_\text{eq}$, $K_\text{eq}$, of the unit cell depend on the properties of the surrounding fluid as well as the effective shell properties, according to 
\begin{subequations} \label{-1}
\begin{align}
\rho_\text{eq}  &= (1-f)\rho + f \rho_\text{eff},
\\
K_\text{eq} &= \Big( (1-f)K^{-1} +fK_\text{eff}^{-1} \Big)^{-1} .
\end{align}
\end{subequations}
Here $f=\pi a^2/d^2$ is the shell volume fraction in the unit cell, where $d$ is the cylinder spacing as well as the side length of the unit cell. 
Since the required density  from TA is $\rho_\text{eq} = \rho$, it follows that the shell effective density is also constant, $\rho_\text{eff} = \rho$. 
The  effective bulk modulus of the shell necessary to achieve the  equivalent value from TA is 
\beq{Kcell}
{K_\text{eff}} = \Big( K^{-1} +\big(K_\text{eq}^{-1} - K^{-1} \big) f^{-1} \Big)^{-1} .
\eeq
\begin{figure}[h!]    
\centering
 \includegraphics[width=3.3in]{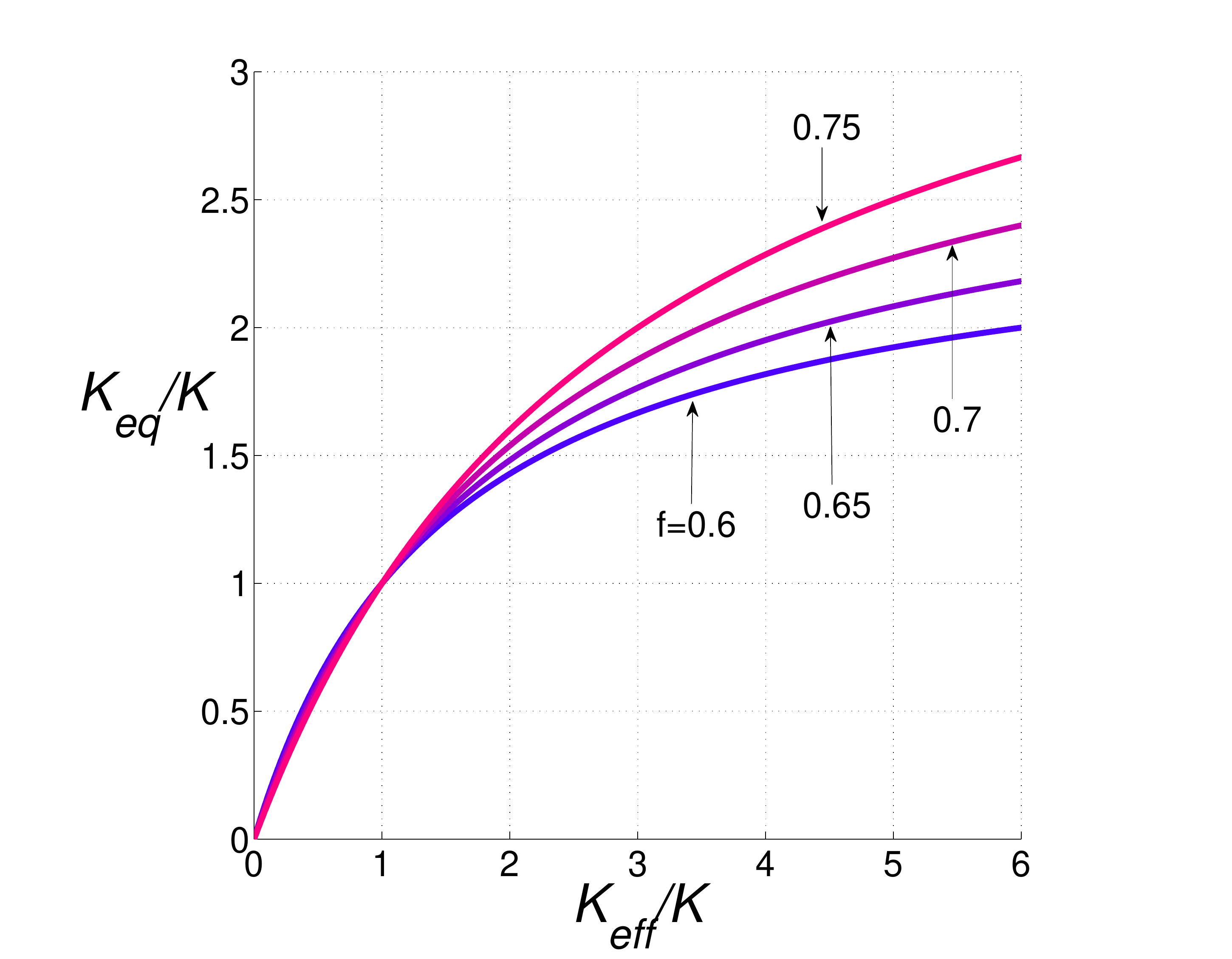}  
\caption{Equivalent bulk modulus of the unit cell $K_\text{eq}$ as a function of the effective bulk modulus of the tuned shell $K_\text{eff}$ for several filling fractions.  (Color online)} 
\label{f4}  
\end{figure}

The equivalent  bulk modulus of the unit cell is significantly affected by the surrounding fluid.  {With the exception of $n=0$, all in-plane modes produce no volume change and hence do not change the effective bulk modulus of the unit cell.  No significant volume altering modes were observed in the frequency range considered.}  Shells of radius $a=1$ cm with a relatively tight packing of $d=2.2a$ yields a filling fraction of $f=0.65$. In this case, in order to have the effective quasi-static bulk modulus of the unit cell $K_\text{eq}=2K$, the effective bulk modulus of the shell-springs-mass system must be $K_\text{eff} = 4.33K$, see Figure~\ref{f4}.


The proposed array contains 7 by 7 unit cells of size $d=2.2a$ with $a=1$ cm giving a  lens side length of  $L=15.4$ cm. The central square element is left empty, requiring 48 cylinders.  This was considered the minimal number necessary to provide  {both a reliable and  an accurate} gradient index effect.  The spacing was chosen to reduce the overall dimension of the lens as much as possible, without making the filling fraction unduly large.  {Inter-cylinder spacing in the fabricated lens was controlled from the two  ends using preformed holders, see Figure \ref{f8} below.} 

Figure~\ref{f5}  shows the discretized values for  the 
equivalent stiffness of each unit cell $K_\text{eq}$ as determined from Figure~\ref{f2} by spatial averaging. 
The  effective properties of the shells are obtained from equation~\eqref{Kcell} with $f=0.65$, using the required equivalent stiffness of each unit cell $K_\text{eq}$ in Figure~\ref{f2}. As noted above, this means that  effective properties of the shells must be more extreme than those implied by the mapping alone. 
The effective density of each shell is tuned to water. 
\begin{figure}[h!]    
\centering
 \includegraphics[width=3.2in]{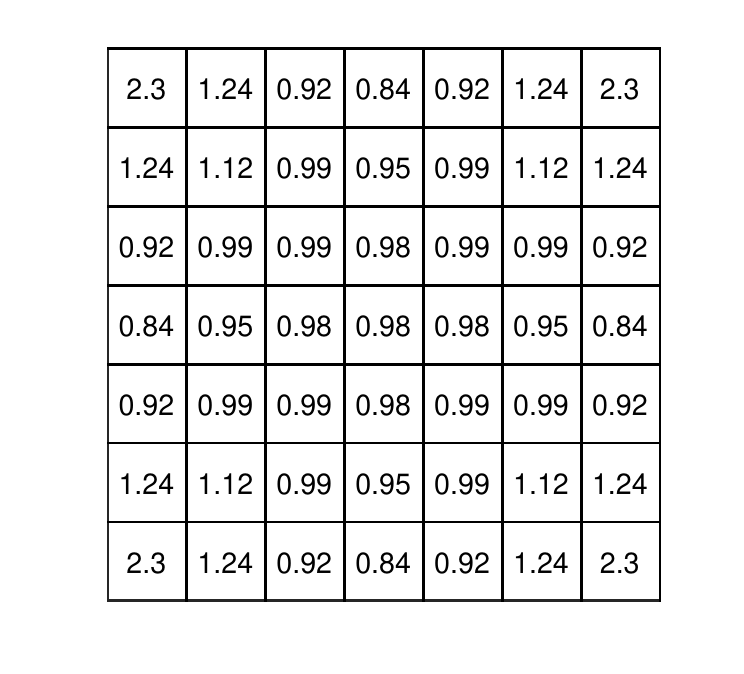}  
\caption{The spatial distribution of the equivalent bulk modulus $K_\text{eq}$ in the 7x7 array.
{The central element is absent (i.e.\ water) in the constructed design.}} \label{f5}  
\end{figure}

\subsection{Design using available cylindrical tubes}

The three primary design criteria were: 
1) that the shells are readily available, 
2) the effective density of each shell {approximately} matches water and 
3) the effect is apparent  in the designated frequency range of interest: near 20 to 25 kHz. 
{The shells must be sub-wavelength in dimension. Furthermore 
all shells are required to have nearly the same outer diameter;
therefore,  the common outer diameter of 0.5 inches is selected as practical.} 
Fixing this outer dimension leaves two parameters: the shell material and its relative thickness $h/a$.
The range of effective properties as a function of both the shell material and thickness are succinctly summarized in Figure~\ref{f1}.  

Several features are apparent from the chart in Figure \ref{f1}.  First, it is clear that the ten materials considered provide a  comprehensive range  of effective properties.  For each material, the effective properties are approximately linear functions of the shell  thickness  for thin shells $(\frac ha \ll 1)$, with some curvature at larger values of $\frac ha$. The present design requires shells with effective density equal to that of water, which restricts  values of $\frac ha$ to those near the  vertical dotted line. Table~\ref{table1} summarizes the properties of available shells which have nearly the same density as water, but varying effective bulk moduli.
\begin{table}  [h!]
\begin{center}
\begin{tabular}{ l  c  c  c  c  c  c  c  c }  
\hline \hline
Material & OD (in) & $h$ (in) & $h/a$ & $\rho_\text{eff}$ & $K_\text{eff}$ &   label \\ \hline 
\text{PVC} 			& 0.54 	& 0.088 & 0.33 	& 0.71 & 0.36 &     1 \\   
\text{ABS} 			& 0.5 	& 0.125 & 0.5 	& 0.98 & 0.52 &     2 \\   
\text{Acrylic} 		& 0.5 	& 0.125 & 0.5 	& 0.89 & 0.68 &   3 \\ 
\text{Polycarbonate} & 0.5 	& 0.125 & 0.5 	& 0.90 & 0.77 &     4 \\ 
\text{Brass} 		& 0.5 	& 0.14 	& 0.056	& 0.93 & 1.63 &     5 \\
\text{Brass} 		& 0.5 	& 0.02 	& 0.08 	& 1.31 & 2.38 &     6 \\ 
\text{Copper} 		& 0.625	& 0.028 & 0.09 	& 1.50 & 2.74 &   7 \\  
\text{Aluminum} 	& 0.5 	& 0.035 & 0.14 	& 0.71 & 2.78 &   8 \\  
\text{Aluminum} 	& 0.5 	& 0.049 & 0.20 	& 0.97 & 4.13 &   9 \\  
\text{Aluminum} 	& 0.5 	& 0.065 & 0.26 	& 1.24 & 5.88 &  10 \\
\hline \hline
\end{tabular}
\caption {Readily available shells (i.e. tubes and pipes) with different effective bulk moduli that have  effective density close to that of water. All properties are normalized to water.} \label{table1}
\end{center}
\end{table}

\begin{figure}[h!]    
\centering
\includegraphics[width=2.7in]{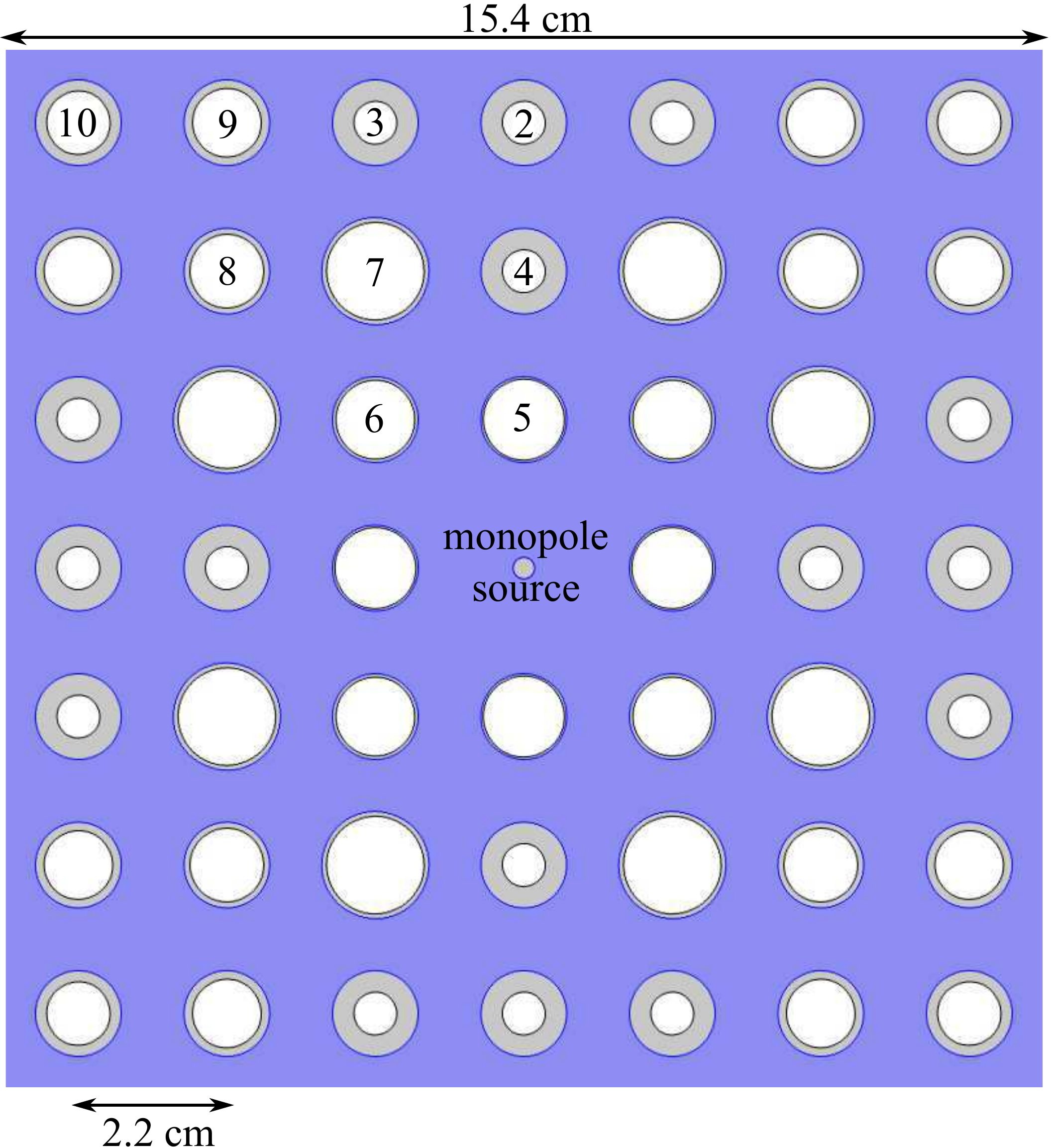} 
\caption{The 7x7 array of various empty shells. The numbers correspond to the index of each shell shown in Table~\ref{table1}. The  actual thicknesses of the individual shells are indicated.  (Color online)}  
\label{f6} 
\end{figure}

For the final design we considered only commercially available tubes made from a variety of materials with standard values of radius and thickness. As Figure~\ref{f1} illustrates, this provides a surprisingly wide range of possible properties, with the added advantage of allowing us  to fabricate the lens with minimal effort and cost.   Based on the available candidates from Table~\ref{table1} we selected nine different shells as shown in Figure~\ref{f6} for the fabricated lens. 


\begin{figure*}[htbp]    
\centering
\subfloat{  \includegraphics[width=1.8in]{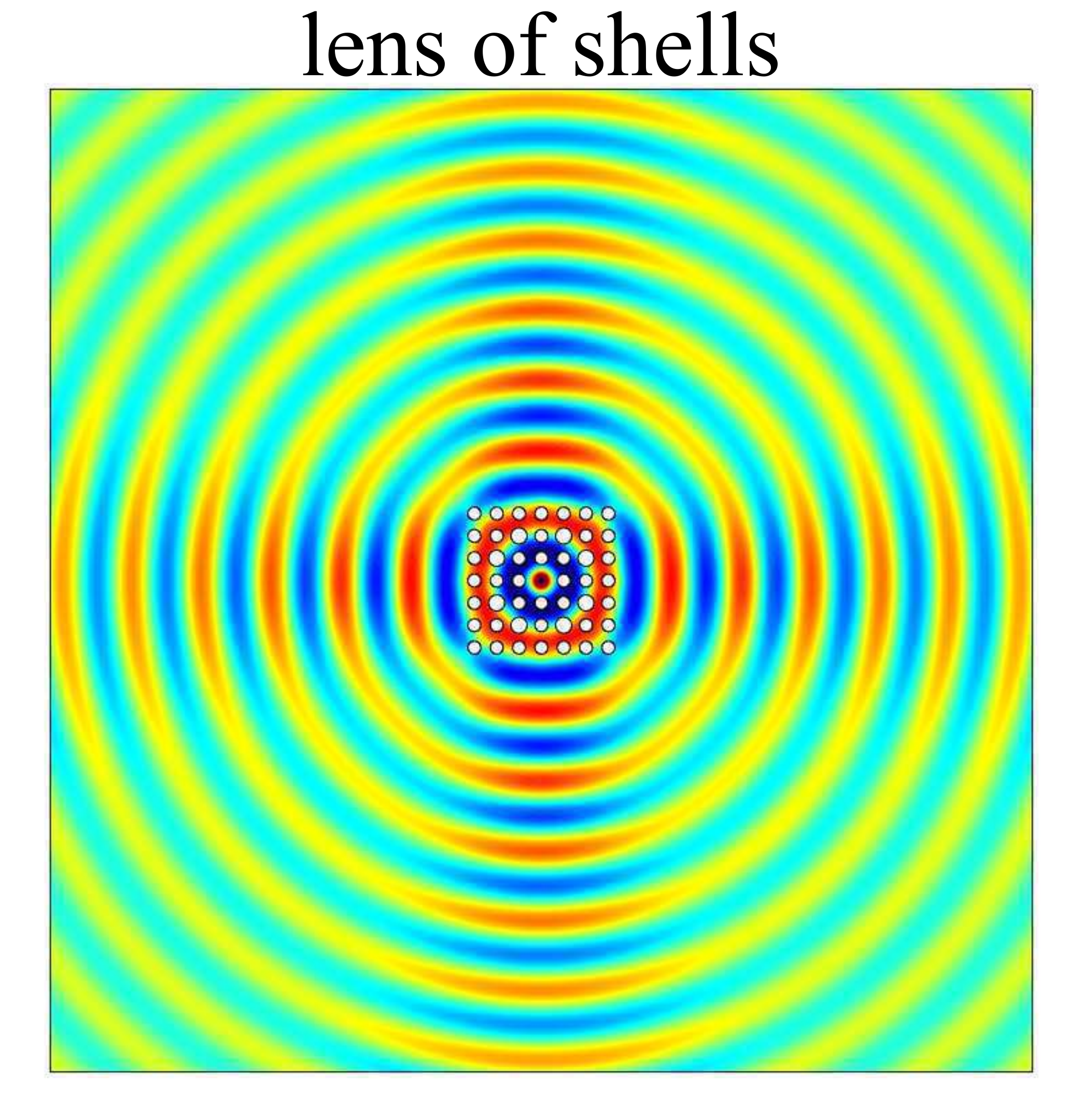}} 
\subfloat{  \includegraphics[width=1.8in]{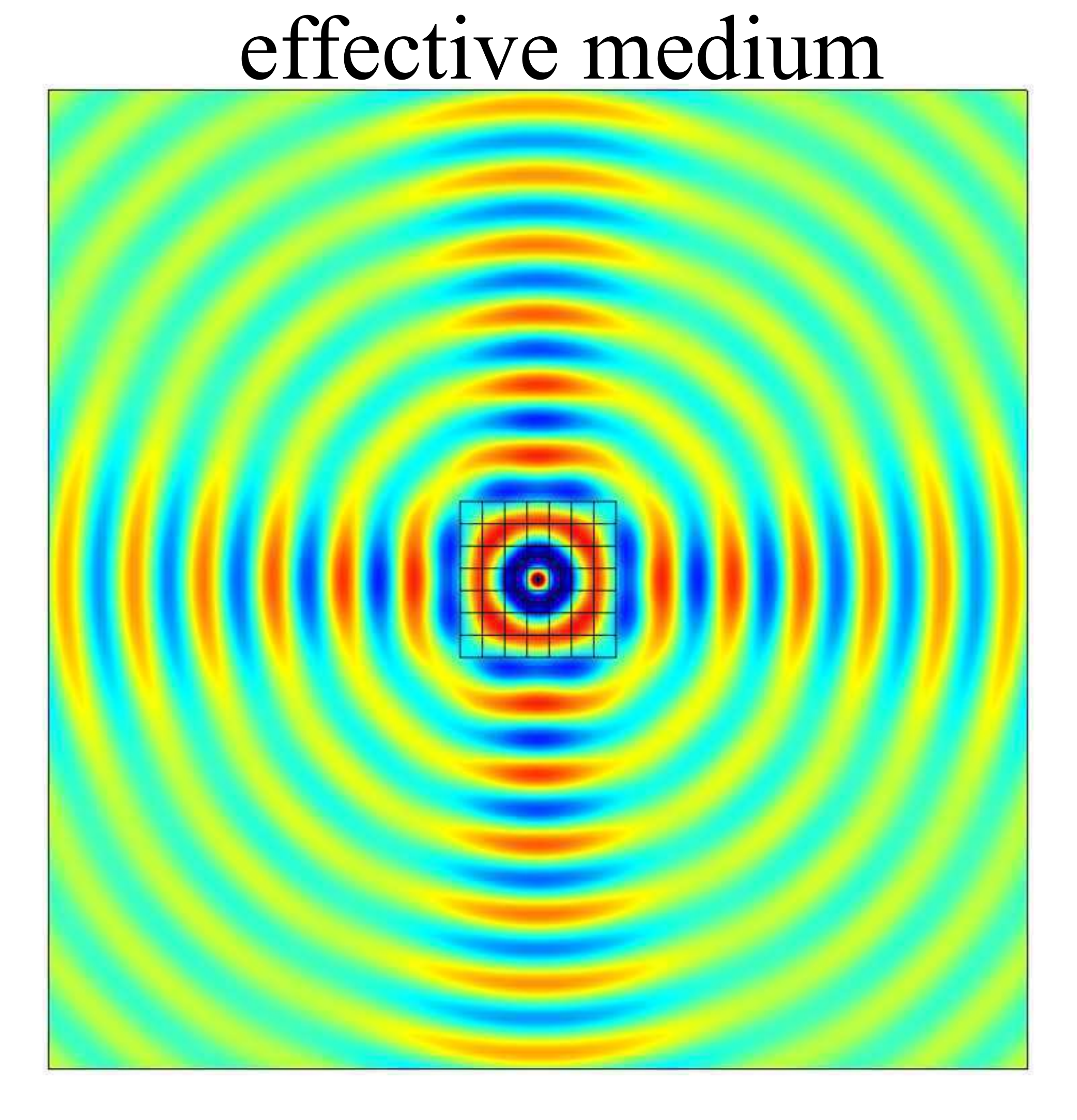}} 
\subfloat{  \includegraphics[width=1.8in]{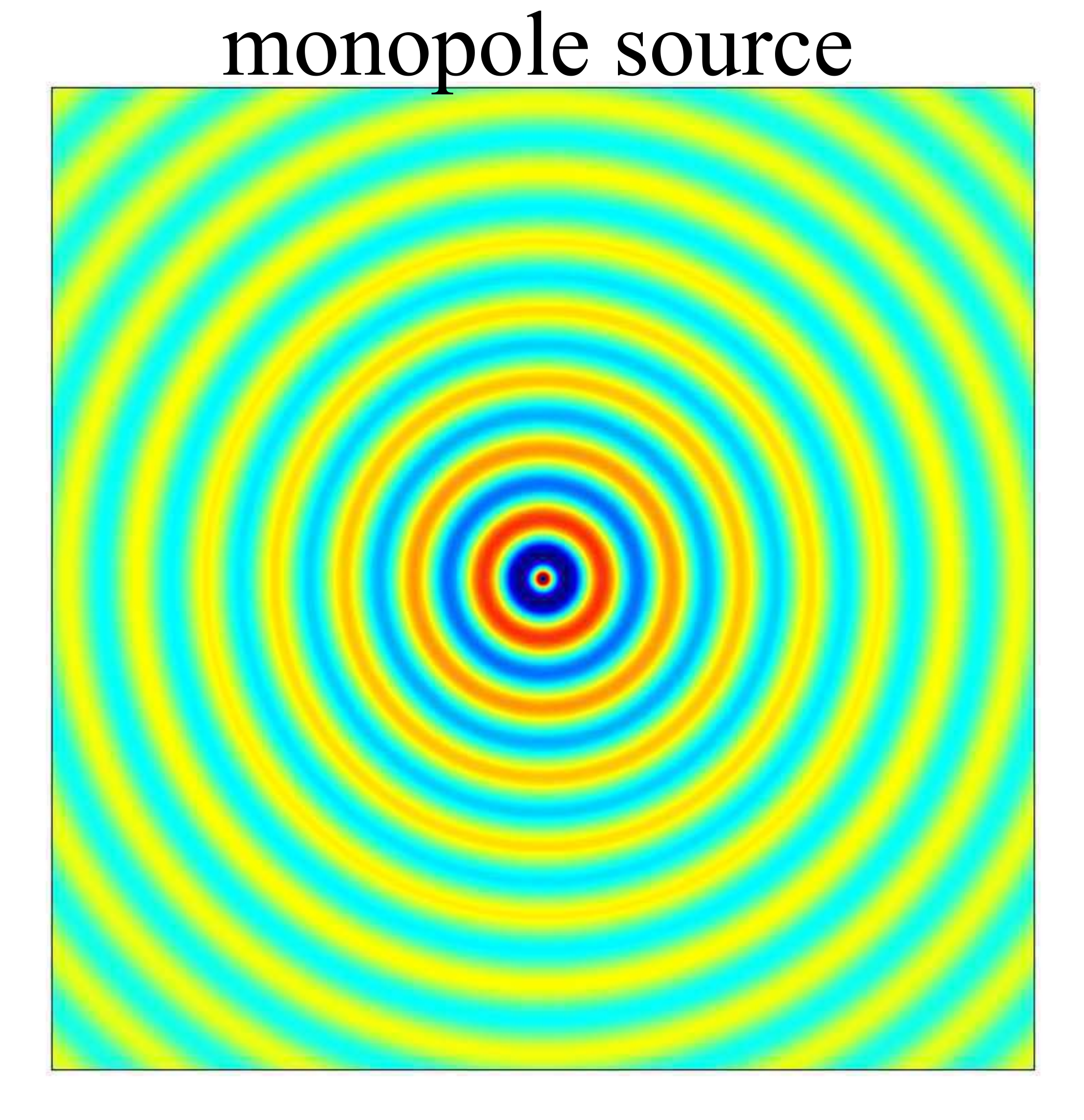}} 
\\
\subfloat{  \includegraphics[width=1.8in]{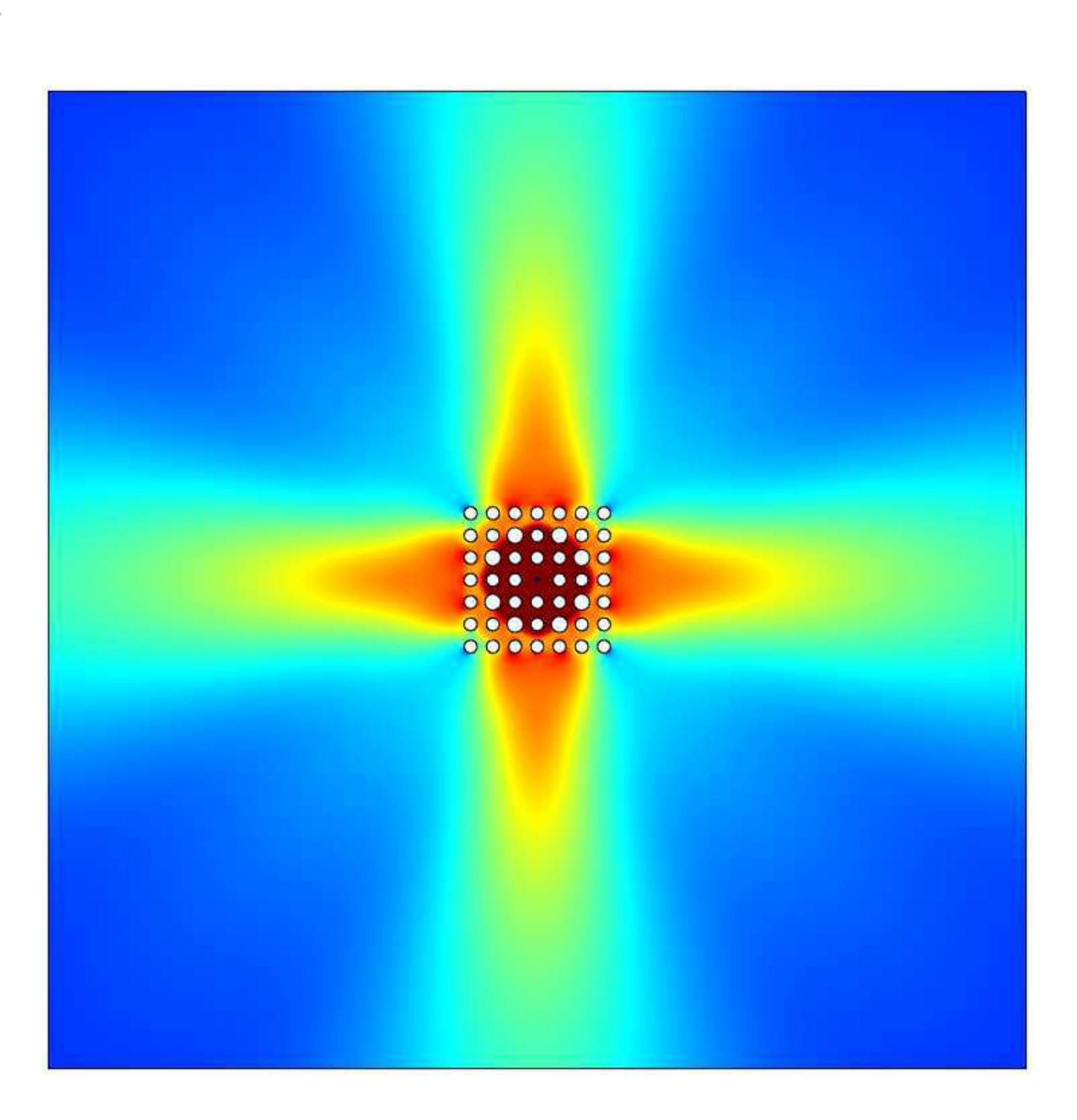}} 
\subfloat{  \includegraphics[width=1.8in]{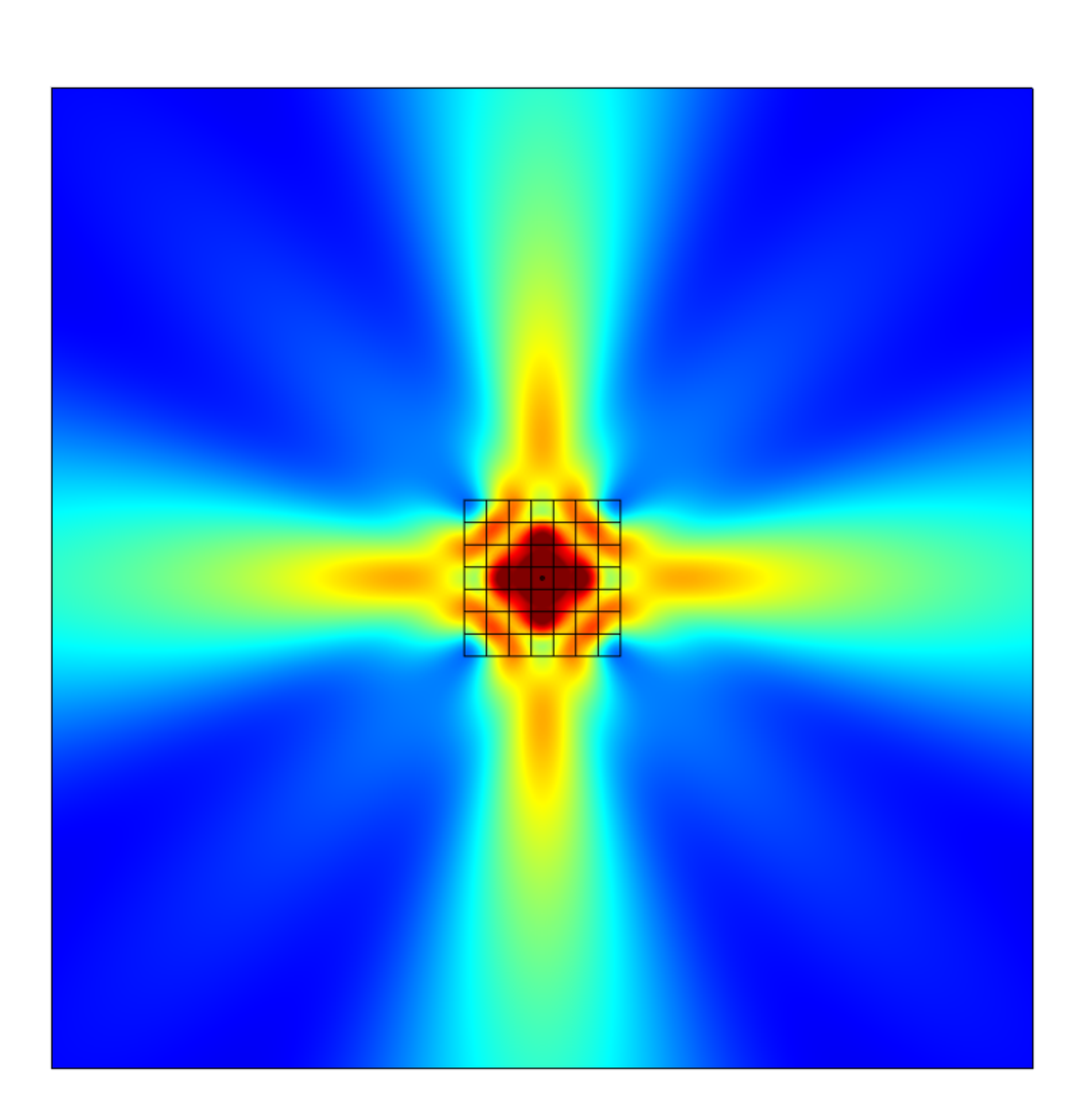}} 
\subfloat{  \includegraphics[width=1.8in]{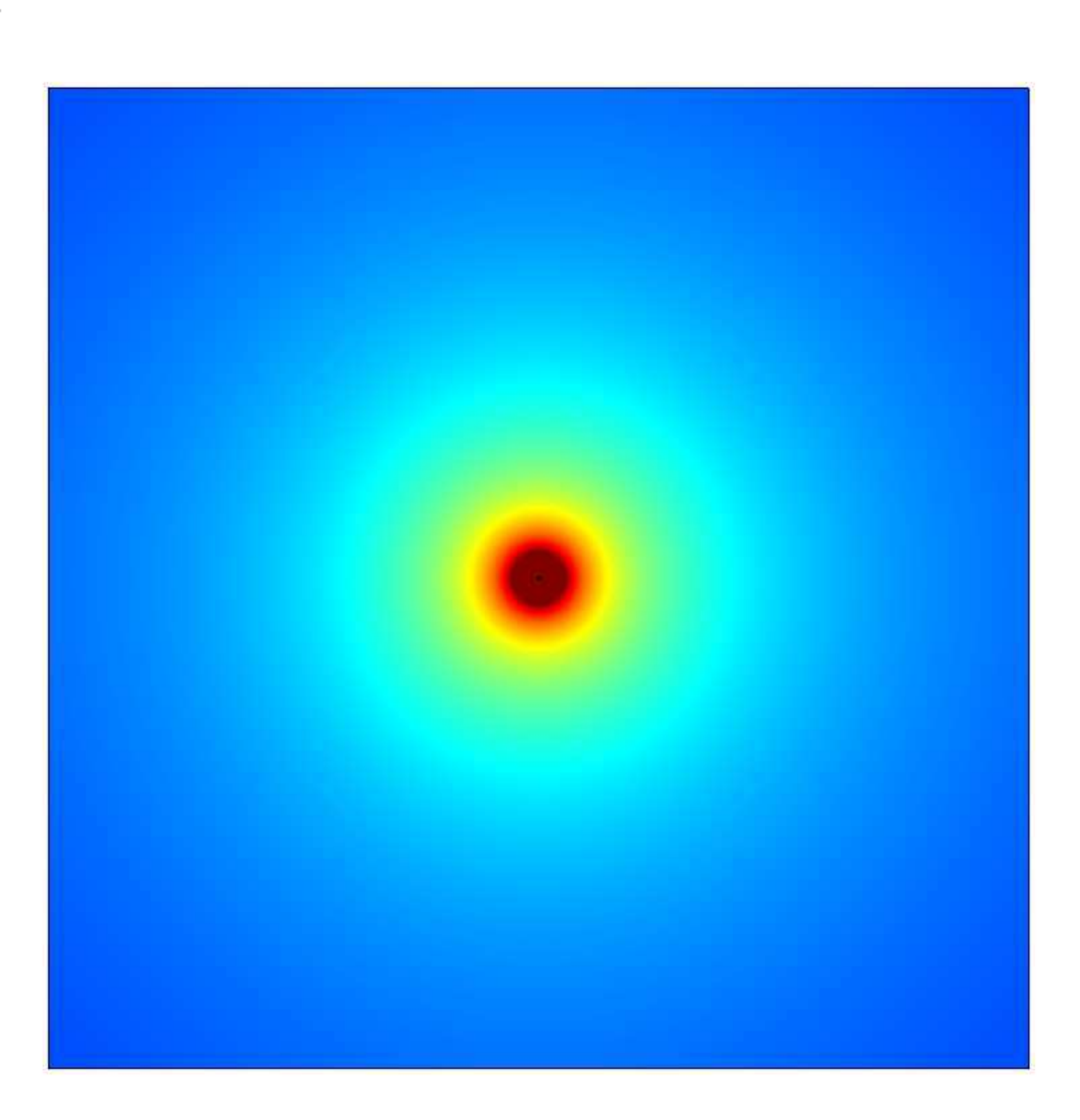}} 
\caption{The left, middle and right columns show the simulated results for the lens of Figure~\ref{f6}, the same square region with the effective acoustic medium from the exact mapping in each unit cell,  and the source without the lens, respectively. The top row is  the total pressure  and the bottom row shows the absolute pressure field for a monopole source at 22 kHz.
 (Color online)} \label{f7} 
\end{figure*}

\subsection{Simulations} 
The total pressure field of the cylindrical-to-square lens made of elastic shells was obtained by numerical computation using COMSOL. 
Figure~\ref{f7} shows a simulation for a monopole source of frequency 22 kHz in the center of the lens. Also shown for comparison are the pressure fields for the lens with the unit cells replaced by the effective acoustic medium  and the free field radiation of the monopole source.

The simulations indicate that the cylindrical-to-plane wave lens made from the distribution of nine distinct empty shells performs very well as compared the optimal case of each unit cell having the prescribed effective acoustic properties directly from the conformal mapping. It is also evident that the transmission is enhanced in the four principal directions.  

The use of empty shells in water acoustics opens up the likelihood of exciting flexural resonances. This is not normally a concern when dealing with isolated shells because the flexural waves are subsonic in speed and hence do not radiate.  The present design places the shells in close proximity, leading to the possibility of coherent flexural wave interaction,   which can lead to strong scattering.   This effect can have positive or negative consequences, depending on one's immediate goal.  In the present situation the shells are of varying  thickness  and comprised of different materials, with the result that the flexural resonances are spread over many frequencies, which decreases the possibility for coherent interaction.  In particular, we note that no such coherent effects were observed in the experiments (see next Section).  Related and surprising  constructive interference effects resulting from  coherent interaction of flexural waves in closely packed arrays of shells in water are described elsewhere \cite{Titovich2015c}.

\section{Experimental results}   \label{sec4}

\begin{figure}[h!]
	\centering
		\includegraphics[width = 2.5in]{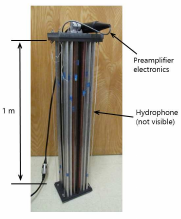} 
	\caption{Constructed device that was tested at Lake Travis Test Station of the Applied Research Laboratories (ARL) at The University of Texas at Austin. The A48 hydrophone is within the lens and the preamplifying electronics are visible. Those electronics are located approximately 1 meter above the lens during the test. The ABS clamping plates and monofilament tensioning elements are visible as are the differing materials of the lens, indicated by changes in cylinder color.  (Color online) }
	\label{f8}
\end{figure}

The device pictured in Figure~\ref{f8} was fabricated to validate the cylindrical-to-plane wave lens design. The device tested has the cylinder positions, radii, and material properties provided in Figure~\ref{f6} and Table~\ref{table1}.   Although the model presented Section \ref{sec3} is strictly two-dimensional (2D), which would suggest experimental validation using a 2D water waveguide, the test facilities available to the authors required a three dimensional (3D) test configuration. Details of the configuration and rationale for their selection are provided here.

The as-tested lens is constructed from cylindrical rods  1~m in height and sealed at either end with urethane end-caps to prevent water intrusion. The cylinders are clamped between 2 cm thick Acrylonitrile-Butadiene-Styrene (ABS) plates using tensioned monofilament. The plates were machined to precisely locate the top and bottom of the cylinders in the positions dictated by the design. Note that an added benefit of the urethane end-caps is that they provide some level of vibration isolation between the end plates and the cylinders. One key challenge to accurately measure the performance predicted in Section \ref{sec3} was to minimize the effects of the finite height of the lens and thus observe its 2D response. The associated practical difficulty encountered was in the selection and placement of the appropriate acoustic source. Validation of the lens design implies the need for an axis-symmetric source pressure along the vertical axis in the 3D lens, but no such source was available to the authors nor could one be easily constructed. Acoustic reciprocity,  described below, was invoked to resolve this difficulty.

Reciprocity is a fundamental principal of quiescent acoustic media, first fully described for acoustics by Rayleigh \cite{ Strutt01111871}; it states that the interchange of source and receiver will lead to the same measured acoustic field if the environment is un-perturbed. Specifically, if one excites acoustic waves at some point, \emph{A}, then “the resulting velocity potential at a second point, \emph{B}, is the same both in magnitude and in phase, as it would have been at \emph{A} had \emph{B} been the source of sound\cite{ Strutt01111871}.”  Applying this principal to the problem at hand, it is possible to replace the axis-symmetric source at the center of the cylindrical-to-plane wave lens with a point receiver and then measure the acoustic field at the center of the lens due to a plane wave incident from a specified radial angle. By varying the angle incidence of the plane wave, it is thus possible to construct the far-field radiation pattern   expected from an axis-symmetric source placed at the center of the lens. The only remaining problem is the generation of plane waves at a specified angle of incidence. This is achieved  using a spherical wave source located   sufficiently far away from the lens  such that the phase of the pressure field impinging on the lens aperture has  variations  less than 1$^\circ$ across the entire frequency band of interest. For the lens geometry and frequencies considered, this can be achieved by   at least 10 m    separation  between the spherical source and the lens.
\begin{figure}[h!]
	\centering
		\includegraphics[width=3.in]{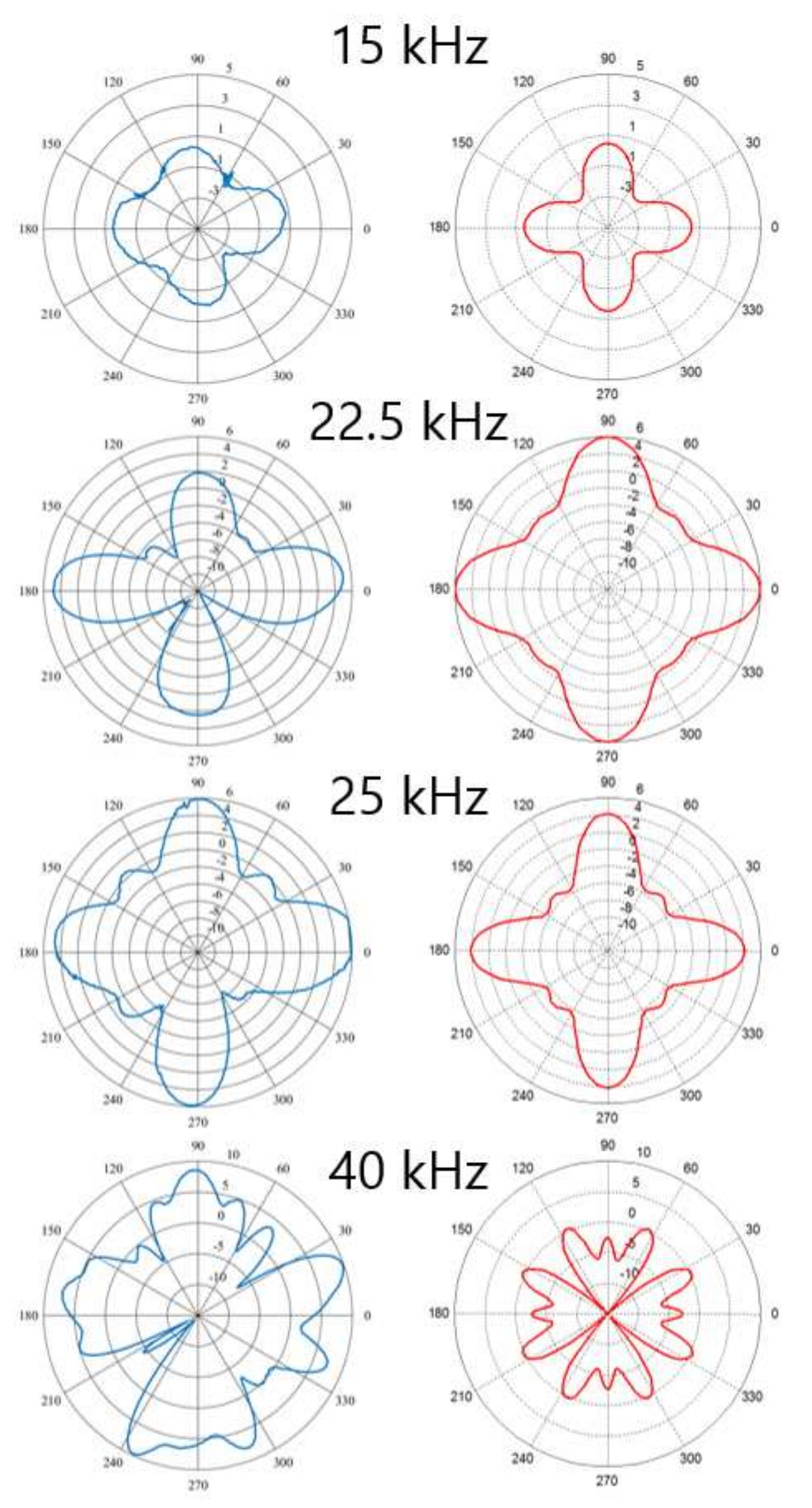}  
	\caption{Angle-dependent gain within the cylindrical-to-plane wave lens at frequencies   15, 22.5, 25, and 40 kHz. The left column presents experimentally obtained angle-dependent gain for a hydrophone within the lens and a plane wave incident from the indicated angle. The right column shows FEM (COMSOL)  calculation of far-field radiation beam pattern for a source in the interior of the lens. All plots show gain (radial coordinate) on the same scale and angle range from 0-360$^\circ$.  (Color online)}
	\label{f9}
\end{figure}

The experiment was conducted at the Lake Travis Test Station (LTTS) of the Applied Research Laboratories (ARL) at The University of Texas at Austin. An A48 hydrophone and associated pre-amplification electronics, which was fabricated and calibrated by the Underwater Sound Reference Division (USRD) of the Naval Undersea Warfare Center (NUWC), was located at the center of the lens. This hydrophone has less than 1 dB of variation across the entire frequency range of interest for this experiment, which was 15--40 kHz. The acoustic source was an omni-directional ITC-1032 fabricated by Channel Technologies Group. The source and lens with internal hydrophone were submerged to a depth of 5.5 m with a separation distance of 10 m. The lens was attached to a column capable of angular rotation through 360$^\circ$. The source is then driven with 2 ms tone bursts from 15--40 kHz at 2.5 kHz intervals and the time-series voltage output from the hydrophone was collected from 0-360$^\circ$ at approximately 0.5$^\circ$  intervals using a sampling frequency of 512 kHz. This process was then repeated for the hydrophone without the lens as a reference and referred to as the bare hydrophone case.

\begin{figure}[htbp]
	\centering
		\includegraphics[width=3.in]{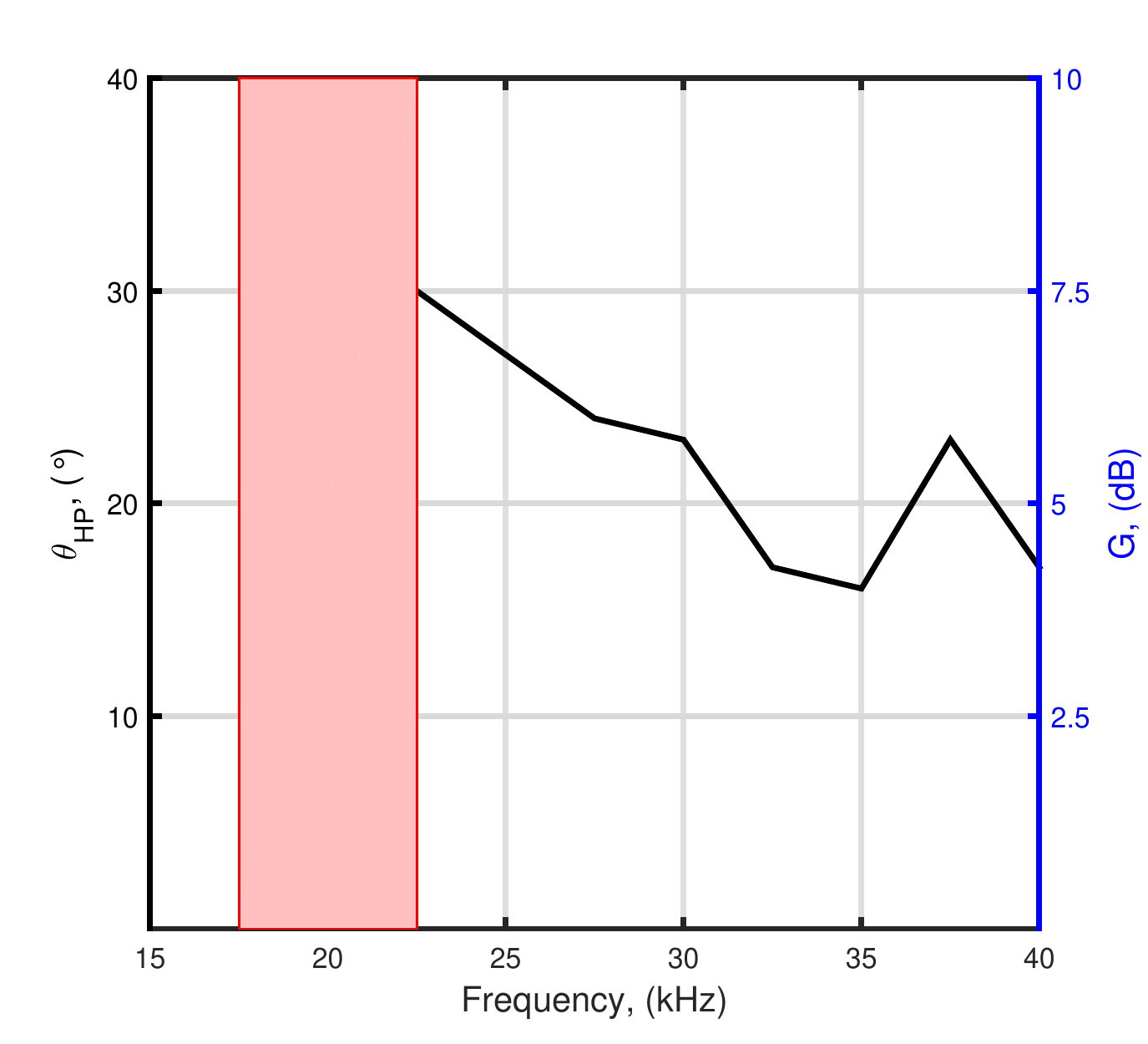} 
	\caption{Half-power beam width (solid line) and average on-axis gain (dot-dash line) of all four lobes of the cylindrical-to-plane wave lens. No beam pattern variations greater than 3 dB were observed for frequencies below 22.5 kHz, and thus no information is provided for $\theta_{HP}<22.5^\circ$. The shaded region (17.5–-22.5 kHz) denotes a flexural tube resonance regime predicted in \cite{Titovich2015c}. Broadband gain and narrow beamwidth is apparent over the entire range of frequencies inspected.  (Color online) }
	\label{f10}
\end{figure}

Representative results from the series of experiments conducted on the cylindrical-to-plane wave lens are summarized in Figures~\ref{f9} and \ref{f10}. The results were obtained by performing post-processing of the time-series data output from the hydrophone,  described next. For each angle and frequency, the steady-state portion of the tone-burst is identified through inspection of the time-domain voltage signal and a time gate is set so that only the steady state portion is considered. The magnitude of the signal at each frequency and angle combination is then found by averaging the magnitude of the complex envelope of the received voltage signal during its steady-state response. This process is carried out for the both measurement configurations (hydrophone in the lens and bare hydrophone). The frequency- and angle-dependent gain is then calculated as $G\left(f,\theta\right) = 20*\mathrm{log}_{10}\left[\lvert V_\mathrm{lens} \left(f,\theta\right)\rvert / \lvert V_\mathrm{no lens} \left(f,\theta\right)\rvert \right]$. Representative polar plots for $f = $ 15, 22.5, 25, and 40 kHz from experimental data and 2D finite element models are shown in Figure~\ref{f9}. Agreement between model and measurement for the both gain and angular dependence (beam pattern) match very well, with the location of the main lobes observed at 4$^\circ$, 91$^\circ$, 176$^\circ$, and 268$^\circ$ on average across all frequencies inspected (with the exception of the 20 kHz case as described below). Unexpected variations in beam pattern between predicted and measured performance are likely owing to imperfections in the constructed device. One very important observation of this data is the broadband performance of this metamaterial lens. The broadband nature of the response is clearly demonstrated by the results provided in Figure~\ref{f10}, which shows the measured half-power beam width (-3 dB points) and on-axis gain averaged at across all four main lobes. The data clearly show that the as-tested lens provides broadband on-axis gain and beam-widths ranging from approximately 15$^\circ$  -– 30$^\circ$ for frequencies  from 22.5 -– 40 kHz, respectively. Finally, it is important to note that the red shaded region in Figures~\ref{f10}  indicates a regime of flexural tube resonances 
\cite{Titovich2015c} where the lens behavior was significantly degraded. This experiment provides clear validation of the broadband impedance matched lensing effect provided by hollow cylinder metamaterial elements.

\section{Conclusion}   \label{sec5}

	The results of  this paper  have shown the practical potential of using cylindrical elastic shells as elements in acoustic metamaterial devices.  
	The demonstration test device considered is a cylindrical-to-plane wave structure for which the required element properties are determined from  transformation acoustics. 
			The size and material composition of the elements in the square array are chosen based on availability of shells, minimizing fabrication difficulties.  The device has the added advantage that is neutrally buoyant by virtue of the transformation acoustics design. 
			Simulations indicated the operating  frequency response of the final design would display a surprisingly broadband effect, which is verified in the experimental findings.   The underwater measurements show effective conversion of the monopolar source to quadropolar radiation  over an  octave band (20 to 40 kHz) with positive gain in the desired directions, all   despite the minimal number of elements used.    These features have been demonstrated for the first time in a water-based acoustic lens device.
Future research will consider other device designs using cylindrical shell passive AMM elements.

\appendix*   
\section{Circle to square mapping}   \label{app}
The {Schwartz-Christoffel} conformal transformation of the unit disk to a square has been   used previously for lens design   via transformation optics \cite{Schmiele10} and transformation acoustics \cite{Layman11}. 
Here we provide a simpler form of the transformation than that given in   \cite{Schmiele10,Layman11}.   

Our objective is a transformation from the plane of the unit circle, defined by the complex variable $\gamma$, to the plane containing the mapped square, defined by the complex variable
$s$ (for "square").  

We first map  the interior of the unit circle to the upper half plane of the variable  $z$ through a bilinear transformation  as $z=i(1-w)/(1+w)$ with $w= \gamma e^{i\pi/4}$. 
The mapping that takes the upper half of the  $z$-plane  to the $s$-plane containing the square is a special case of the more general mapping known for 
mapping to polygons. Thus, consider $f(z) =A  + Bg(z)$ where 
\begin{equation}\label{1-5}
g(z) =  \int_0^{z} \prod\limits_{i=1}^3 (\zeta-\zeta_i)^{-1/2} \dd \zeta .
\end{equation}
Taking $\zeta_1=0$, $\zeta_2=1$, $\zeta_3=-1$, we find 
$g(z) =  \sqrt{2}F\big(\sqrt{z+1}; 1/{\sqrt{2}}\big)$ 
where $F$ is the incomplete elliptic integral of the first kind.  The parameters  $A$ and $B$ are found by setting $f(0)\rightarrow 1-i$, $f(-1) \rightarrow -1-i$, and using   
$F \big(0;\frac 1{\sqrt{2}} \big)=0$,  
$F\big(1;\frac 1{\sqrt{2}} \big) = K\big(\frac 1{\sqrt{2}}\big) = \frac{1}{4\sqrt{\pi}}
\Gamma^2\Big(\frac 14 \Big)$,
where $K(k)$ is the complete elliptic integral of the first kind and $\Gamma(n)$ is the gamma function. 
Hence, in terms of the original $\gamma$-plane containing the unit circle
\begin{equation}\label{1-13}
s(\gamma) = \frac{2}{K(\frac 1{\sqrt{2}})} F\bigg(\sqrt{1+i\frac{1-\gamma e^{i\pi/4}}{1+\gamma e^{i\pi/4}}};\frac 1{\sqrt{2}} \bigg) -1-i .
\end{equation}\label{1-14}
Equation \eqref{1-13} and its inverse   map the boundary points in the N, S, E, W, NE, NW, SE and SW directions in the circle and square plane to one another.

The density and bulk modulus are functions of the derivative of the mapping function.  The derivative of \eqref{1-13} is found  from $f '(z) = \sqrt{2}/(K(\frac 1{\sqrt{2}})\sqrt{z(z^2-1)})$ and  $z'(w)=-2i/(1+w)^2$, which  gives $s ' (\gamma)$.  
%
Hence, for  $\gamma = r e^{i\theta}$, $0\leq\theta<2\pi$ and $0\leq r\leq 1$, 
\begin{equation} \label{1-15a}
|s '(r e^{i\theta})|=\frac{2}{K(\frac 1{\sqrt{2}})} \big(r^8 + 1 +2r^4\cos 4\theta\big)^{-1/4}.
\end{equation}
The inverse mapping from the square, $s$ coordinate, to the circle, $\gamma$ coordinate is given by eq.\ (27) of Titovich and Norris \cite{Titovich14a}.

\acknowledgments        
This work was supported by ONR through MURI Grant No. N00014-13-1-0631 and ULI Grant No. N00014-13-1-0417. Many thanks to Dr. Maria Medeiros of ONR (Code 333) and Dr. Stephen O’Regan of NSWCCD (Code 7220).


\end{document}